\documentclass[preprint,10pt]{sigplanconf}

\usepackage{amsmath}
\usepackage{MnSymbol}%
\usepackage{wasysym}%

\usepackage[textsize=footnotesize,obeyDraft,draft]{todonotes}

\usepackage{tikz}
\usetikzlibrary{arrows,positioning,fit,decorations.pathreplacing,trees,shapes.multipart}
\usepackage{listings}
\usepackage{xspace}
\usepackage{paralist}
\usepackage[group-separator={,}]{siunitx}
\usepackage{xcolor, colortbl}
\usepackage[hyphens]{url}
\usepackage{hyperref}
\begin{document}

\newcommand{\libeacof}{\mbox{\emph{libeacof}}\xspace}
\newcommand{\libeacofbox}[1]{\nodepart{#1} \rotatebox{90}{\small{libeacof}}}
\newcommand{\eacof}{\mbox{\textsc{eacof}}\xspace}
\newcommand{\provider}{Provider\xspace}
\newcommand{\consumer}{Consumer\xspace}
\newcommand{\providers}{Providers\xspace}
\newcommand{\consumers}{Consumers\xspace}
\newcommand{\checkpoint}{Checkpoint\xspace}
\newcommand{\checkpoints}{Checkpoints\xspace}
\newcommand{\centralauthority}{Central Authority\xspace}
\newcommand{\ca}{CA\xspace}
\newcommand{\probe}{Probe\xspace}
\newcommand{\probes}{Probes\xspace}

\newcommand{\functionname}[1]{\mbox{\emph{#1()}}\xspace}
\newcommand{\variablename}[1]{\mbox{\emph{#1}}\xspace}
\newcommand{\structnamenx}[1]{\mbox{\emph{#1}}}
\newcommand{\structname}[1]{\structnamenx{#1}\xspace}
\newcommand{\probet}{\structnamenx{Probe}\xspace}
\newcommand{\probets}{\structnamenx{Probe}s\xspace}
\newcommand{\datarequestt}{\structnamenx{dataRequest\_t}\xspace}
\newcommand{\checkpointt}{\structnamenx{Checkpoint}\xspace}
\newcommand{\checkpointts}{\structnamenx{Checkpoint}s\xspace}

\newcommand{\customparagraph}[1]{\textit{\textbf{#1.}}\xspace}

\setlength{\pdfpageheight}{\paperheight}
\setlength{\pdfpagewidth}{\paperwidth}

\conferenceinfo{CONF 'yy}{Month d--d, 20yy, City, ST, Country} 
\copyrightyear{20yy} 
\copyrightdata{978-1-nnnn-nnnn-n/yy/mm} 
\doi{nnnnnnn.nnnnnnn}




\titlebanner{EACOF}        
\preprintfooter{A Framework for Providing Energy Transparency to enable Energy-Aware Software Development}   

\title{EACOF: A Framework for Providing Energy Transparency to enable Energy-Aware Software Development}

\authorinfo{Hayden Field, Glen Anderson, Kerstin Eder}
           {University of Bristol\\ Department of Computer Science\\
            Merchant Venturers Building\\ Woodland Road\\
            Bristol, BS8 1UB, UK}
           {\{hayden-field.2011, glen.anderson.2011, kerstin.eder\}@bristol.ac.uk}

\maketitle

\begin{abstract}
  Making energy consumption data accessible to software developers is
  an essential step towards energy efficient software engineering.
The presence of various different, bespoke and incompatible, methods
of instrumentation to obtain energy readings is currently limiting the
widespread use of energy data in software development.
This paper presents \eacof, a modular Energy-Aware Computing Framework
that provides a layer of abstraction between sources of energy data
and the applications that exploit them.  \eacof replaces platform
specific instrumentation through two APIs---one accepts input to the
framework while the other provides access to application software.
This allows developers to profile their code for energy consumption in
an easy and portable manner using simple API calls.
We outline the design of our framework and provide details of the API
functionality.  In a use case, where we investigate the impact of data
bit width on the energy consumption of various sorting algorithms, we
demonstrate that the data obtained using \eacof provides interesting,
sometimes counter-intuitive, insights.
All the code is available online under an open source license. {\tt http://github.com/eacof}
\end{abstract}




\section{Introduction}
\label{sec:intro}

Limiting energy use is important in situations such as robotics, portable devices and data centres.
Traditionally, reductions in energy usage have been achieved through improved hardware design, however hardware designers must make conservative assumptions about usage patterns to ensure that their devices remain suitable for a wide range of use cases.
There is a far greater potential for energy saving when the requirements of software are considered~\cite{roy19976}.
This is because the control over the computation ultimately lies within the software and algorithms running on the hardware.

Developers can produce more energy efficient software by implementing their data structures and algorithms appropriately, basing decisions on their knowledge of the needs of a particular application with respect to runtime, space utilisation and energy consumption.
Knowledege of a user's specific usage patterns and priorities provides further opportunity to reduce energy consumption either by passively monitoring a user's interaction with a device and adapting accordingly~\cite{te2013design}, or by providing direct feedback about the energy needed to perform different tasks, enabling users to make informed decisions about their behaviour~\cite{wilke2013comparing}.

In order to leverage this domain specific knowledge to produce more energy efficient software, developers first need to know how much energy is required to execute their code.
Energy usage profiles must be created separately for each targeted platform as it is not possible to generalise from one to another~\cite{pallister2013identifying}.
In some instances it is possible to create these profiles through the use of static analysis~\cite{liqatlopstr2013}; however, in general, dynamic measurement of energy usage is required due to the limited availability and usability of static techniques.
While dynamic measurement techniques are more widely available than static analyses, the overhead of instrumenting code is greatly increased by the low level of abstraction at which energy usage is measured, analogous to requiring machine-specific code to measure execution time.

This paper presents \eacof, our modular \textbf{E}nergy \textbf{A}ware \textbf{CO}mputing \textbf{F}ramework, which provides a layer of abstraction between sources of energy data and the applications that exploit them, allowing developers to profile their code for energy consumption in a simple and portable manner.
This abstraction replaces platform specific instrumentation through the use of two APIs.
The first is used to make energy information available to other software in a portable way while the second is used to access energy information without the need for bespoke instrumentation on each platform.
By separating these two key tasks in a modular manner, our framework is designed to both be easily extensible and encourage the development of maintainable code.

\eacof is designed to be used on a general purpose multi-process Operating System with functional Inter-Process Communication capabilities.
It allows sophisticated data collection methods to provide energy consumption data to applications through a simple API.
The provided data can be as specific as per-process values for individual hardware components.
This energy data may be utilised by a developer during the development process for energy consumption testing.
Alternatively, it may provide the end user with indications about energy consumption at run time or allow the development of applications which adapt based on measured energy consumption.

The rest of this paper is structured as follows.
Existing work upon which \eacof builds is discussed in Section~\ref{sec:background}.
The characteristics of the different sources of energy information that are available to \eacof are considered in Section~\ref{sec:data-sources}.
The various components that make up \eacof, and their interactions, are described in Section~\ref{sec:sys-arch}.
The APIs to allow interaction between components are specified in Section~\ref{sec:interface}.
In Section~\ref{sec:case-study} we present a use case showing how \eacof can be used to analyse sorting algorithms to help a developer to select the most appropriate algorithm or data type for a given task.
Section~\ref{sec:conclusions} concludes and gives an outlook on future work.

\section{Background}
\label{sec:background}

Tools, such as PowerScope~\cite{flinn1999powerscope}, provide functionality to monitor the energy usage of code, while requiring additional hardware for this task.
On Android devices, AppScope~\cite{yoon2012appscope} estimates energy consumption of each process based on a static model; however, this data is not available to the application itself.
The Intel Power Gadget API~\cite{intel:power-gadget-api} allows for the measurement of energy consumption on modern Intel CPUs, although does not provide access to data from other hardware components.
JouleUnit~\cite{wilke2013jouleunit} is designed to be a generic framework, however has a specific focus on testing during development.
The authors of Eprof~\cite{eprof} describe the challenges in attributing the energy usage of hardware to the correct application.

Each of these systems, many of which are designed for mobile devices, uses a bespoke method of instrumentation tailored to a specific source of data.
With \eacof, we provide a standard method of accessing sources of energy data, including these and others, designed primarily for desktop systems.
We also address the stated limitations of each existing source.

\section{Data Source Considerations}
\label{sec:data-sources}

For \eacof to succeed as an abstraction layer for sources of energy information, it is important that its design does not preclude the use of any particular data source since doing so would force developers to use platform specific instrumentation in their applications.
Below we outline the considerations that different data sources place on \eacof.
Section~\ref{sec:providers} shows how \eacof has been designed to meet these considerations.

\customparagraph{Resolution}
The operating system on a laptop might provide a new value for the current charge of the battery once per second, while counters built into a CPU may update thousands or millions of times per second.

\customparagraph{Precision}
The values regarding the current charge of a battery may only be specified to 3 or 4 significant figures within system files, while hardware counters can be significantly more precise.

\customparagraph{Accuracy}
While some data sources take measurements directly from hardware, others utilise a more indirect approach.
Tools such as PowerTOP~\cite{intel:powertop} are able to provide estimates of energy consumption based on usage statistics provided by the operating system, after a calibration period.
Because the data is not gathered directly from hardware, accuracy needs to be taken into account when making use of the provided values.

\customparagraph{Probes}
A \probet is a means of defining one or more hardware sources of energy data (devices).
Energy data can cover a range of devices.
This can range from individual CPU cores to an entire system, with various steps along the way.
Sometimes, a single hardware probe will be attached to multiple devices, causing the energy information provided to be an aggregate of a number of devices.

\customparagraph{Units}
The units in which data is collected can vary from one source to another---one source might provide energy data in Joules while another provides power data in Watts.

\customparagraph{Temporal Continuity}
Some sources of data may temporarily be unavailable when attempts are being made to use them.
For example, plugging a laptop in to charge will eliminate the ability to use the discharge rate of the battery as a measure of whole system energy usage.

\customparagraph{Proliferation}
As we place greater demands on systems to be energy-efficient, we will see an increasing number of sources of energy data built directly into the hardware platforms we use.
Many of these will have a higher resolution and accuracy than current data sources.


\section{System Architecture}
\label{sec:sys-arch}

In this section we describe the components of our framework and how they interact with each other.

\providers (\S~\ref{sec:providers}) abstract the details of a data source, making it available to higher level components in a portable manner.
\consumers (\S~\ref{sec:consumers}) of energy data are able to access the information made available by \providers.
The \centralauthority (\S~\ref{sec:ca}) marshals data between the \providers and \consumers as well as providing several additional services.
Each of these components can be developed and compiled independently and run as separate system processes, providing a simple method of modularly extending the functionality of \eacof.
Our device classification system (\S~\ref{sec:device-classification}) provides a common method for the components of \eacof to describe the hardware being monitored.
We have utilised various strategies (\S~\ref{sec:dealing-with-overheads}) to minimise the overhead incurred from use of \eacof.

A library, \libeacof, is provided to abstract the mechanisms underlying the communication between components from the programmer, providing procedures that can be readily integrated into portable application code.
The two APIs provided by \libeacof are described in Section~\ref{sec:interface}.

\subsection{Data Providers}
\label{sec:providers}

\begin{figure}[width=\linewidth]
  \centering
  \resizebox{0.75\linewidth}{!}{
    \begin{tikzpicture}[auto]
      \tikzset{
        rectangleNode/.style={rectangle,draw=black,thick, inner sep=1em, minimum size=2em,text centered},
        splitRectangleNode/.style={rectangle,rectangle split, rectangle split horizontal, rectangle split parts=2, draw=black,thick, inner sep=0.5em, text centered},
        bigNode/.style={rectangle,draw=black,very thick, inner sep=2em, minimum size=3em,text centered},
        arrow/.style={->, >=latex', shorten >=1pt, thick}
      }

      \node[rectangleNode, text width=5ex] (cpu) {CPU};
      \node[rectangleNode, below=0.6cm of cpu, text width=5ex] (hdd) {HDD};

      \node[splitRectangleNode, right=0.5cm of cpu] (provider1) {\nodepart[text width=15ex]{one} CPU monitor \libeacofbox{two}};
      \node[splitRectangleNode, right=0.5cm of hdd] (provider2) {\nodepart[text width=15ex]{one} HDD monitor \libeacofbox{two}};
      \node[fit=(provider1)(provider2)](providerGroup){};

      \draw[arrow] (cpu.east) -- (provider1.west);
      \draw[arrow] (hdd.east) -- (provider2.west);

      \node[fit=(provider1)(provider2)](providerGroup){};

      \coordinate[yshift=-1ex] (brace point) at (provider2.south);
      
      \draw[line width=1pt,decorate,decoration={amplitude=7pt,brace,mirror}]
      (brace point -| provider2.west)
        -- node[anchor=base,yshift=-4ex] {Providers}
      (brace point -| provider2.text split);

      \draw[line width=1pt,decorate,decoration={amplitude=7pt,brace,mirror}]
      (brace point -| hdd.west)
        -- node[anchor=base,yshift=-4ex] {Data Sources}
      (brace point -| hdd.east);

      \node[rectangleNode, text width=12ex, right=of providerGroup] (ca) {Central Authority};
      \draw[arrow] (provider1.east) -- (ca.west);
      \draw[arrow] (provider2.east) -- (ca.west);

    \end{tikzpicture}
  }
  \caption{Central Authority with multiple Providers}
  \kern-1em
  \label{fig:providers}
\end{figure}
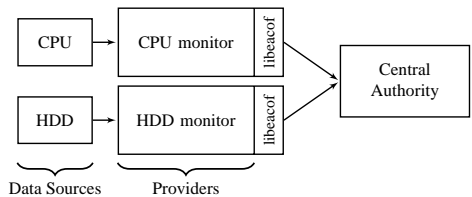

A \emph{\provider} is software that provides energy information to \eacof.
The \centralauthority acts as a single point of contact between all components in \eacof.
Figure~\ref{fig:providers} shows how a \provider acts as a wrapper for a source of energy information, collecting data in a platform specific way before passing this data to the \centralauthority in a portable format.
It is relieved of the need to separate out per-process data by the \centralauthority, which undertakes this task on behalf of all \providers.
Since \providers act as a wrapper for a data source, they are designed to handle possible variations between existing and future sources of data.
The following describes our solutions to considerations about data sources:

\customparagraph{Resolution}
\providers push their data to the \centralauthority rather than waiting for it to be requested.
This ensures that sampling is performed in the most appropriate way for a given data source.
For example, if a data source updates in a non-linear manner, the \provider can ensure that the \centralauthority always has the latest data no matter what the gap between updates.

\customparagraph{Precision}
Use of \structnamenx{double}s for energy consumption data provides precision for decimal values with up to 15 significant digits, as specified in IEEE 754.
This is the greatest level of precision obtainable with portable data types.

\customparagraph{Accuracy}
Since a \provider may encapsulate any data source, \eacof provides no guarantees about the accuracy of a \provider's data.
It is up to the developer of a \provider to ensure the data output is as accurate as possible and the developers of \consumers to cope with data of varying accuracy.

\customparagraph{Probes}
\providers define \probets that specify their functionality.
Section~\ref{sec:device-classification} describes in detail the method by which devices may be defined within \eacof.

\customparagraph{Units}
It is a role of a \provider to convert the value gathered from its data source into Joules before passing it on to the \centralauthority.
This ensures that \consumers only need to handle a single unit, Joules.

\customparagraph{Temporal Continuity}
A \provider lets other components know when its capabilities change, either temporarily or permanently.
Section~\ref{sec:provider-api} describes the API we have specified to allow this functionality.

\customparagraph{Proliferation}
New sources of energy information can be supported through the development of new \providers.
These new \providers can be integrated into \eacof without the need to modify the \centralauthority or pre-existing \providers due to the modular framework design.
Our device classification system (\S~\ref{sec:device-classification}) ensures that \consumers can automatically make use of new \providers without alteration.

\subsection{Data Consumers}
\label{sec:consumers}
\label{sec:checkpoints}

\begin{figure}[width=\linewidth*0.5]
  \centering
  \resizebox{0.5\linewidth}{!}{
    \begin{tikzpicture}[auto]
      \tikzset{
        rectangleNode/.style={rectangle,draw=black,thick, inner sep=1em, minimum size=2em,text centered},
        splitRectangleNode/.style={rectangle,rectangle split, rectangle split horizontal, rectangle split parts=2, draw=black,thick, inner sep=0.5em, text centered},
        bigNode/.style={rectangle,draw=black,very thick, inner sep=2em, minimum size=3em,text centered},
        arrow/.style={->, >=latex', shorten >=1pt, thick}
      }
      \node[rectangleNode, text width=12ex] (ca) {Central Authority};
      \node[splitRectangleNode, right=of ca] (consumer2) {\libeacofbox{one} \nodepart[text width=14ex]{two} Video Player};
      \node[splitRectangleNode, above=0.2cm of consumer2] (consumer1) {\libeacofbox{one} \nodepart[text width=14ex]{two} Web Browser};
      \node[splitRectangleNode, below=0.2cm of consumer2] (consumer3) {\libeacofbox{one} \nodepart[text width=14ex]{two} Background File Sync};
      \draw[arrow] (ca.east) -- (consumer1.west);
      \draw[arrow] (ca.east) -- (consumer2.west);
      \draw[arrow] (ca.east) -- (consumer3.west);

      \coordinate[yshift=-1ex] (brace point) at (consumer3.south);
      
      \draw[line width=1pt,decorate,decoration={amplitude=7pt,brace,mirror}]
      (brace point -| consumer3.text split)
        -- node[anchor=base,yshift=-4ex] {Consumers}
      (brace point -| consumer3.east);
    \end{tikzpicture}
  }
  \caption{Central Authority with multiple Consumers}
  \kern-1em
  \label{fig:consumers}
\end{figure}
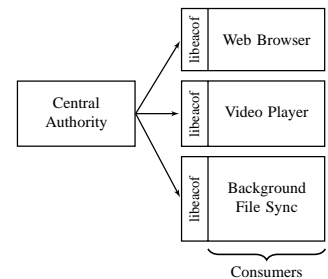

A \emph{\consumer} is an application that makes use of energy data provided by \eacof.
Figure~\ref{fig:consumers} shows how the \centralauthority is able to connect to a number of \consumers simultaneously.
Each \consumer has no awareness of the \providers that feed energy information into \eacof---as far as it is concerned, all the information accessed is provided by the \centralauthority itself.

\consumers determine their energy consumption through the use of \checkpoints.
\emph{\checkpoints} are a means of sampling the energy used by a combination of one or more devices between multiple points in time.
They are added into an application's code through the use of the \consumer API (\S~\ref{sec:consumer-api}), allowing the energy consumption of sections of code to be determined.
There are three key points in the timeline of a \checkpoint: creating, sampling, and deleting.

To use a \checkpoint, it must first be created, setting a point in time to use as a baseline for the energy being measured.
When creating a \checkpoint, the devices and processes that are to be measured must be specified.
A \checkpoint can be used to measure the \consumer itself or all running processes.
Upon creation of a \checkpoint, the \consumer must communicate with the \centralauthority to acquire a unique identifier for the \checkpoint and to ensure that the requested devices can be monitored.

To obtain energy readings, a \checkpoint must be sampled.
When a \consumer samples a \checkpoint, the \centralauthority will provide the number of Joules used by the specified processes on associated devices since the \checkpoint was last sampled by the same \consumer.
Creating a \checkpoint is deemed to be the first sample point.

It is possible for a single \consumer to have multiple active \checkpoints, each monitoring one or more devices.
Likewise, it is possible for multiple \consumers to be monitoring the same device.

\subsection{Central Authority}
\label{sec:ca}

\begin{figure}[width=\linewidth]
  \centering
  \resizebox{\linewidth}{!}{
    \begin{tikzpicture}[auto]
      \tikzset{
        rectangleNode/.style={rectangle,draw=black,thick, inner sep=1em, minimum size=2em,text centered},
        splitRectangleNode/.style={rectangle,rectangle split, rectangle split horizontal, rectangle split parts=2, draw=black,thick, inner sep=0.5em, text centered},
        bigNode/.style={rectangle,draw=black,very thick, inner sep=2em, minimum size=3em,text centered},
        arrow/.style={->, >=latex', shorten >=1pt, thick}
      }

      \node[rectangleNode, text width=5ex] (cpu) {CPU};
      \node[rectangleNode, below=0.6cm of cpu, text width=5ex] (hdd) {HDD};

      \node[splitRectangleNode, right=0.5cm of cpu] (provider1) {\nodepart[text width=15ex]{one} CPU monitor \libeacofbox{two}};
      \node[splitRectangleNode, right=0.5cm of hdd] (provider2) {\nodepart[text width=15ex]{one} HDD monitor \libeacofbox{two}};
      \node[fit=(provider1)(provider2)](providerGroup){};

      \draw[arrow] (cpu.east) -- (provider1.west);
      \draw[arrow] (hdd.east) -- (provider2.west);

      \node[fit=(provider1)(provider2)](providerGroup){};

      \coordinate[yshift=-1ex] (brace point) at (provider2.south);
      
      \draw[line width=1pt,decorate,decoration={amplitude=7pt,brace,mirror}]
      (brace point -| provider2.west)
        -- node[anchor=base,yshift=-4ex] {Providers}
      (brace point -| provider2.text split);

      \draw[line width=1pt,decorate,decoration={amplitude=7pt,brace,mirror}]
      (brace point -| hdd.west)
        -- node[anchor=base,yshift=-4ex] {Data Sources}
      (brace point -| hdd.east);

      \node[rectangleNode, text width=12ex, right=of providerGroup] (ca) {Central Authority};
      \draw[arrow] (provider1.east) -- (ca.west);
      \draw[arrow] (provider2.east) -- (ca.west);

      \node[splitRectangleNode, right=of ca] (consumer2) {\libeacofbox{one} \nodepart[text width=14ex]{two} Video Player};
      \node[splitRectangleNode, above=0.2cm of consumer2] (consumer1) {\libeacofbox{one} \nodepart[text width=14ex]{two} Web Browser};
      \node[splitRectangleNode, below=0.2cm of consumer2] (consumer3) {\libeacofbox{one} \nodepart[text width=14ex]{two} Background File Sync};
      \draw[arrow] (ca.east) -- (consumer1.west);
      \draw[arrow] (ca.east) -- (consumer2.west);
      \draw[arrow] (ca.east) -- (consumer3.west);

      \coordinate[yshift=-1ex] (brace point) at (consumer3.south);
      
      \draw[line width=1pt,decorate,decoration={amplitude=7pt,brace,mirror}]
      (brace point -| consumer3.text split)
        -- node[anchor=base,yshift=-4ex] {Consumers}
      (brace point -| consumer3.east);
    \end{tikzpicture}
  }
  \caption{EACOF System Architecture}
  \label{fig:central-authority}
\end{figure}
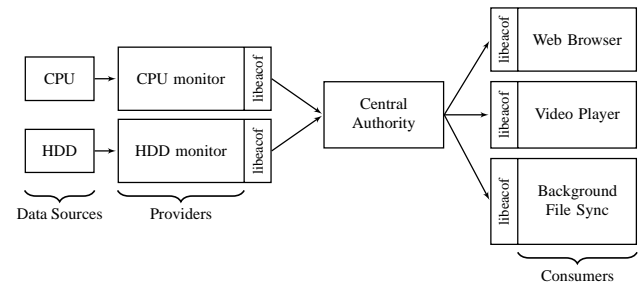

The \emph{\centralauthority} acts as a single point of contact between all other components in \eacof.
Figure~\ref{fig:central-authority} shows how these components fit together, with data flowing from a data source all the way through to a \consumer, as directed by the arrows.

By communicating with all other components, the \centralauthority is able to perform operations that \providers and \consumers would be unable to complete alone.
For example, let there be two \probes defined by \providers---one measuring both the CPU and Memory, the other measuring only the Memory.
If a \consumer then defines a \checkpoint to track energy consumption of the CPU, the \centralauthority is able to derive this requested information by reading data from both \probes and calculating the difference.


The \centralauthority is also able to keep track of recent readings from multiple \providers and extrapolate from them to estimate future energy consumption even if the resolution of available \providers is low.
A \consumer could perform this task, however, performing it in a central location makes it easier to integrate the full capabilities of \eacof into an application.

\subsection{Device Classification}
\label{sec:device-classification}

\begin{figure}
\newlength{\socketTextWidth}
\settowidth{\socketTextWidth}{Socket N}
\newlength{\coreTextWidth}
\settowidth{\coreTextWidth}{Core N}
\newlength{\hddTextWidth}
\settowidth{\hddTextWidth}{HDD N}

  \centering
  \resizebox{0.5\linewidth}{!}{
  \tikzstyle{every node}=[rectangle,draw,thick,anchor=west]
  \begin{tikzpicture}[
  grandchild/.style={grow=down,xshift=1em,anchor=west,
    edge from parent path={(\tikzparentnode.south) |- (\tikzchildnode.west)}},
  first/.style={level distance=6ex},
  second/.style={level distance=12ex},
  third/.style={level distance=18ex},
  fourth/.style={level distance=24ex},
  fifth/.style={level distance=30ex},
  sixth/.style={level distance=36ex},
  seventh/.style={level distance=42ex},
  eighth/.style={level distance=48ex},
  level 1/.style={sibling distance=5em}]
    \node
      {System}
    [edge from parent fork down]
    child{node {CPUs}
      child[grandchild,first] {node[text width=9.4ex,align=center] {Socket 0}
        child[grandchild,first] {node[text width=7.8ex,align=center] {Core 0}}
        child[grandchild,second] {node[text width=7.8ex,align=center] {Core 1}}
        child[grandchild,third] {node[text width=7.8ex,align=center]{\vphantom{Core}\ldots}} 
        child[grandchild,fourth] {node[text width=7.8ex,align=center]{Core N}}
      }
      child[grandchild,sixth,text width=9.4ex,align=center] {node [text width=\socketTextWidth]{Socket 1}}
      child[grandchild,seventh] {node [text width=9.4ex,align=center] {\phantom{C}\ldots\phantom{C}}}
      child[grandchild,eighth,text width=9.4ex,align=center] {node {Socket N}}
    }
    child [missing] {}
    child {node{HDDs}
      child[grandchild,first] {node[text width=8.1ex,align=center] {HDD 0}}
      child[grandchild,second] {node[text width=8.1ex,align=center] {HDD 1}}
      child[grandchild,third] {node[text width=8.1ex,align=center] {\vphantom{HDD}\ldots}}
      child[grandchild,fourth] {node[text width=8.1ex,align=center] {HDD N}}
    }
    child {node [align=center] {\vphantom{Socket}\ldots}};
  \end{tikzpicture}
  }
  \caption{Classification of a subset of devices}
  \kern-1em
  \label{fig:device-classification}
\end{figure}
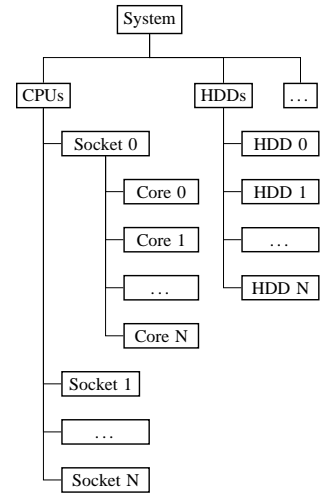

The definition of a particular device is key to a clear common understanding between components within \eacof.
A device classification system must be precise enough that data received by \consumers meets their expectations while also being flexible enough to allow switching of \probes with equivalent functionality in a manner transparent to \consumers.


Figure~\ref{fig:device-classification} shows a subset of the classification system used by \eacof.
While starting at a very high System level, our classification system provides scope to zoom in on specific components.
Energy can be provided and consumed for devices at any level of this classification tree, for example at the whole System level, for all CPUs, or for a single core of a particular CPU.

The classification system sets expectations of \emph{what} is being monitored rather than \emph{how} it is being monitored.
For example, readings from energy counters in the CPU in Socket~0 and the power supply to Socket~0 itself would both be classified as Socket~0.

It is the responsibility of a \provider to specify which devices the \probets it contains represents.
Similarly, it is up to each \consumer to make it known to other components which devices it is interested in obtaining data for.


\subsection{Dealing with Overheads}
\label{sec:dealing-with-overheads}
When running with a single \provider sampling CPU energy consumption at a rate of 50Hz, \eacof increases system power consumption by less than 1 Watt on an 2012 MacbookPro 13" with a 2.5GHz Intel Core i5.
This is lower than the idle power consumption of many common applications~\cite{steigerwald2012energy}.

When \eacof is used to profile code during development, these overheads are of little concern as they will not be incurred by the final application.
However, when \eacof is an integral part of the application it is vital that developers are able to mitigate these overheads when necessary.
This can be achieved by sampling fewer devices less frequently or by instrumenting only the most critical sections of code.

The overheads incurred when using \eacof must always be weighed against the benefits.
While static analysis can deliver results without runtime overhead, there are limits to what can be determined in this manner.
In such situations, the only alternative is to use a method of dynamic analysis such as \eacof, where the inherent overheads are necessary to achieve the desired outcome.

The \centralauthority plays a key role in minimising the overheads incurred when using \eacof by reducing the number of communication channels that need to be established.
More formally, let the number of \consumers be $C$ and the number of \providers be $P$.
In a naive configuration, where \consumers communicate directly to \providers, $O(C*P)$ communication channels are required.
This is reduced to $O(C+P)$ by introducing the \centralauthority.
In adding an extra step between the \provider and \consumer, latency is increased, however this design prevents a poorly designed \provider blocking a \consumer request for a long period---it is easier to ensure predictable latency within a single \centralauthority than within many \providers.




\begin{figure*}[width=\linewidth]
  \begin{lstlisting}[language=c,breaklines=true,breakatwhitespace=true,basicstyle=\small,morekeywords={Probe,DeviceCount},numbers=left,rulecolor=\color{black},frame=single]
int createProbe(Probe **newprobe, DeviceCount dcount, ...);
int deleteProbe(Probe **probe);
int activateProbe(Probe *probe);
int deactivateProbe(Probe *probe);
int addSample(Probe *probe, double joules);
  \end{lstlisting}
  \kern-1em
  \caption{\provider API Function Prototypes}
  \label{fig:provider-api}
\end{figure*}

\begin{figure*}[width=\linewidth]
  \begin{lstlisting}[language=c,breaklines=true,breakatwhitespace=true,basicstyle=\small,morekeywords={Checkpoint,ProcessSpecifier,DeviceCount},numbers=left,rulecolor=\color{black},frame=single]
int setCheckpoint(Checkpoint **newcheckpoint, ProcessSpecifier pspec, DeviceCount dcount, ...);
int sampleCheckpoint(Checkpoint *checkpoint, double *joules);
int deleteCheckpoint(Checkpoint *checkpoint);
  \end{lstlisting}
  \kern-1em
  \caption{\consumer API Function Prototypes}
  \kern-1em
  \label{fig:consumer-api}
\end{figure*}

\section{Interface}
\label{sec:interface}

To allow components of \eacof to interact with each other we have developed two APIs---one for \providers, the other for \consumers.
The \provider API (\S~\ref{sec:provider-api}) is used to extend the framework with additional sources of energy information independently of the development and compilation of the framework itself.
The \consumer API (\S~\ref{sec:consumer-api}) is used by software developers wishing to exploit the gathered information, either for the purposes of profiling or as an integral part of the design of their application---perhaps to provide energy awareness to end users.
These APIs have been designed to work together and may be used simultaneously by a single application.

Our implementation of the API is available in an online repository at: {\tt http://github.com/eacof}

\subsection{\provider API}
\label{sec:provider-api}





A \probet is an abstract data type representing a capability of a \provider and can be thought of as a virtual description of a physical \probe used to monitor one or more devices.
The API functions shown in Figure~\ref{fig:provider-api} are used to create, manipulate and delete \probets.

\functionname{createProbe} is a variadic function used to specify the configuration for a \probet.
\libeacof will allocate a new \probet and populate it with the devices provided by the caller.
A unique identifier for the new \probet is acquired from the \centralauthority and the address of the new \probet is stored in the location pointed to by \variablename{newprobe}.

While a \probet is inactive to begin with, \functionname{activateProbe} and \functionname{deactivateProbe} may be used to toggle whether it is active.
Once finished with, \functionname{deleteProbe} will delete a \probet and stop further attempts at trying to access the data it provides.


\functionname{addSample} provides the amount of energy used, in Joules, by the devices associated with the supplied \probet.
The sample is the amount of energy used since the latter of \begin{inparaenum}[\itshape a\upshape)]
\item the last point at which the \probet was activated; or
\item the last point at which a sample was added.
\end{inparaenum}

The return value of all API functions in this section and the following is an error code.
This allows users of the API to know when a requested operation failed, making it clear when alternative action, such as retrying the function call or entering a non-energy-aware state, should be taken.


\subsection{\consumer API}
\label{sec:consumer-api}

A \checkpointt is an abstract data type representing a \checkpoint as described in Section~\ref{sec:checkpoints}.
The API functions shown in Figure~\ref{fig:consumer-api} are used to create, sample and delete \checkpointts.

\functionname{setCheckpoint} is a variadic function used to specify a set of devices to be monitored along with the processes they should be monitored for, creating a \checkpoint with the given configuration.
\variablename{pspec} should either be \variablename{ALL} or \variablename{SELF} depending on whether the caller wishes to monitor all running processes or itself respectively.

\functionname{sampleCheckpoint} is used to query the amount of energy in Joules used by the process-device combination specified by \variablename{checkpoint} since the latter of \begin{inparaenum}[\itshape a\upshape)]
\item the last call to \functionname{sampleCheckpoint}; or
\item the point at which the \checkpoint was created.
\end{inparaenum}

\functionname{deleteCheckpoint} is used to signal that the \checkpointt \variablename{checkpoint} is no longer required, deleting it so it may not be used again until recreated.

\begin{table*}[width=\linewidth]
	\centering
	\resizebox{\linewidth}{!}{
	\begin{tabular}{ | r r || r | r | r || r | r | r || r | r | r || r | r | r | }
		\hline
		& & \multicolumn{12}{ c | }{Data Type} \\
		\cline{3-14}
		& & \multicolumn{3}{ c || }{uint8\_t} & \multicolumn{3}{ c || }{uint16\_t} & \multicolumn{3}{ c || }{uint32\_t} & \multicolumn{3}{ c | }{uint64\_t} \\
		\cline{3-14}
		& & \multicolumn{1}{c|}{Total} & \multicolumn{1}{c|}{Total} & \multicolumn{1}{c||}{Average} & \multicolumn{1}{c|}{Total} & \multicolumn{1}{c|}{Total} & \multicolumn{1}{c||}{Average} & \multicolumn{1}{c|}{Total} &  \multicolumn{1}{c|}{Total} & \multicolumn{1}{c||}{Average} & \multicolumn{1}{c|}{Total} & \multicolumn{1}{c|}{Total} & \multicolumn{1}{c|}{Average} \\
		& & \multicolumn{1}{c|}{Time} & \multicolumn{1}{c|}{Energy} & \multicolumn{1}{c||}{Power} & \multicolumn{1}{c|}{Time} & \multicolumn{1}{c|}{Energy} & \multicolumn{1}{c||}{Power} & \multicolumn{1}{c|}{Time} &  \multicolumn{1}{c|}{Energy} & \multicolumn{1}{c||}{Power} & \multicolumn{1}{c|}{Time} & \multicolumn{1}{c|}{Energy} & \multicolumn{1}{c|}{Power} \\
		Algorithm & Num Elements & \multicolumn{1}{c|}{(s)} & \multicolumn{1}{c|}{(J)} & \multicolumn{1}{c||}{(W)} & \multicolumn{1}{c|}{(s)} & \multicolumn{1}{c|}{(J)} & \multicolumn{1}{c||}{(W)} & \multicolumn{1}{c|}{(s)} & \multicolumn{1}{c|}{(J)} & \multicolumn{1}{c||}{(W)} & \multicolumn{1}{c|}{(s)} & \multicolumn{1}{c|}{(J)} & \multicolumn{1}{c|}{(W)} \\
		\hline
		Bubble Sort & \num{50000} & 5.53 & 66.66 & 12.03 & 5.39 & 65.29 & 12.09 & 5.66 & 69.05 & 12.19 & 5.78 & 71.83 & 12.41 \\
		Insertion Sort & \num{200000} & 7.98 & $\blacksquare$102.18 & 12.75 & 7.98 & $\blacksquare$103.00 & 12.85 & 7.46 & $\blacksquare$98.81 & 13.21 & 7.54 & $\blacksquare$105.03 & 13.89 \\
		Quicksort & \num{2000000} & 5.51 & 61.73 & 11.20 & 5.53 & 61.90 & 11.19 & 5.52 & 61.60 & 11.15 & 5.51 & 62.90 & $\bigstar$11.42 \\
		Merge Sort & \num{60000000} & $\bullet$6.06 & $\bullet$72.33 & 11.93 & 6.07 & 72.46 & 11.93 & 6.12 & 75.65 & 12.36 & $\bullet$5.93 & $\bullet$76.98 & $\bigstar$12.98 \\
		qsort & \num{100000000} & $\bullet$5.84 & $\bullet$72.39 & 12.37 & 6.15 & 76.90 & 12.48 & 6.79 & 86.29 & 12.69 & $\bullet$5.69 & $\bullet$73.25 & 12.86 \\
		Counting Sort & \num{200000000} & 0.23 & $\blacklozenge$2.92 & 12.75 & 0.24 & $\blacklozenge$3.16 & 13.23 & 0.25 & $\blacklozenge$3.58 & 14.15 & 0.35 & $\blacklozenge$5.12 & 14.44 \\
		\hline
	\end{tabular}
	}
	\caption{Comparison of the energy required to sort integers of different bit widths}
	\kern-1em
	\label{fig:case-study-results}
\end{table*}

\section{Use Case Example}
\label{sec:case-study}

As an example, this section demonstrates how \eacof can be applied to gain insight into the energy usage of code.
While energy-efficient software design is beyond the primary goals of \eacof, the framework makes energy consumption during computation transparent, so as to enable developers to gain an insight into the energy usage of code.
This use case demonstrates that our framework provides data for developers to make more informed decisions about software energy consumption.
It is not, however, designed to provide an exhaustive demonstration of all the capabilities of \eacof---a demonstration of multi-provider and multi-platform functionality is available in the online repository.

We use a \provider based on the Intel Power Gadget API, with a resolution of 20ms, to measure the energy usage of the CPU in Joules.

Our application is a \consumer designed to sort an array of integers in the range $[0, 255]$.
The same array of numbers is represented using data types containing varying numbers of bits and sorted using a number of standard deterministic sorting algorithms.
\checkpoints are set and sampled directly before and after the sorting occurs.
This setup allows us to examine the difference in energy consumption caused by using different data types to perform a task, a typical problem that a developer may wish to use \eacof to solve.
Since the focus is that of differences between data types, comparisons between rows should not be made for Time or Energy in Table~\ref{fig:case-study-results}.

We ran the code on a 2012 MacbookPro 13" with a 2.5 GHz Intel Core i5 and 8GB RAM running OS X 10.8.4 to gather results.
All code was compiled on the target machine using the vendor's standard compiler with an optimisation level of \textit{-O3}.
The instrumented code used to gather the results is available in an online repository at: {\tt http://github.com/eacof}

Table~\ref{fig:case-study-results} shows the results gained from running our program.
Time is measured in Seconds (s) and Energy is measured in Joules (J).
The average power consumption, energy over time, is measured in Watts (W).
Each algorithm was run with the specified number of elements $200$ times for each data type.
The displayed figures are the calculated means over all $200$ runs.
We keep run time similar by providing each algorithm a different length array as input, so as to allow measurement of algorithms of differing algorithmic complexities.
Counting Sort takes significantly less time per run than the other algorithms because it is memory limited and would require an array containing around 4 billion elements to take a similar amount of time as the others.
The standard deviation of all result values is low, apart from for Insertion Sort where the standard deviation for both Time and Energy is around 10\% of the mean values.

The data collected with \eacof demonstrates that, against common intuition, time and energy consumption are not necessarily directly correlated.
It can be seen in the cells marked with a $\bullet$ in Table~\ref{fig:case-study-results} that with Merge Sort and \functionname{qsort} from \variablename{stdlib}~\cite{kernighan1988c}, sorting 64 bit values takes less time than sorting values with fewer bits, however more energy is consumed in the process.

It can also be seen that the average power consumption when using a single data type varies between algorithms ($\bigstar$).
In a situation where there is a limited power supply, it may be desired to choose an algorithm with lower power consumption even if it means the time taken or energy used is higher.

Another insight that can be gained is that the amount of energy used to perform a sorting task will generally increase as the number of bits in the data type increases.
In our data, marked with a $\blacklozenge$, Counting Sort uses 75\% more energy to sort the same 200 arrays of numbers when they are represented as 64 bit rather than 8 bit values.
Similar increases are not consistent for all algorithms---Insertion Sort broke an otherwise increasing trend when values were represented using a 32 bit data type ($\blacksquare$).

Each of these examples highlight insights that can be gained by having access to the energy consumption information provided by \eacof.
To take a step further, one would want to investigate why Insertion Sort uses less energy when using 32 bit values.
This investigation is beyond the scope of this project, however the issue would not have come up without access to the data.

\section{Conclusions and Future Work}
\label{sec:conclusions}


We have created \eacof, a framework which allows access to information about the energy consumption of software through the use of simple API calls.
This allows the development of code that may portably provide energy transparency to both the end user and developer.
In turn, it enables informed decisions to be made with respect to trade-offs regarding energy consumption.

\eacof is designed to separate the two tasks of collecting and utilising dynamic energy consumption data through the use of \providers and \consumers.
This separation allows for a clear distinction of the two tasks, reducing the development and maintenance overhead required for the successful completion of each task.
In addition, our modular design allows for functionality to improve over time without recompilation of all existing system components.

As demonstrated in our use case (Section~\ref{sec:case-study}), our framework provides the functionality required to obtain values regarding software energy consumption.

We intend to continue developing \eacof and use it to study the improvements in energy-efficiency that can be gained in real-world applications.
We also intend to perform analysis similar to our use case with a range of \providers on a variety of hardware set-ups with respect to other algorithms, data structures and software constructs.
Alongside, we will perform further validation of the figures provided by \eacof to determine how best to obtain accurate data.

\section{Acknowledgments}
The authors would like to acknowledge the contributions of Benji
Barash, Charlie McNeil, James Pedlingham, Tom Ryczanowski and Gary
Noble to the initial development of \eacof.
The research leading to these results has received funding from the
European Union's Seventh Framework Programme (FP7/2007-2013) under
grant agreement no 611004, project ICT-Energy.







\bibliographystyle{abbrvnat}
\bibliography{eacof}





\end{document}